\documentclass[fleqn,twoside,twocolumn,nofootinbib]{revtex4} 
\usepackage[sec]{ujp}

\begin{document}
\title[Investigation of hadron multiplicities and hadron yield ratios in heavy ion collisions]
{Investigation of hadron multiplicities and hadron yield ratios in heavy ion collisions}

\author{D.R. Oliinychenko}
\affiliation{Bogoliubov Laboratory of Theoretical Physics}
\address{6, Joliot-Curie Str., JINR, Dubna 141980, Russia}
\email{dimafopf@gmail.com}

\author{K.A. Bugaev}
\affiliation{\bitp}
\address{\bitpaddr}
\email{bugaev@th.physik.uni-frankfurt.de}

\author{A.S. Sorin}
\affiliation{Bogoliubov Laboratory of Theoretical Physics}
\address{JINR, Joliot-Curie str. 6, 141980 Dubna, Russia}
\email{Sorin@theor.jinr.ru}

\udk{539.12}  \razd{\seci}

\setcounter{page}{211}%
\maketitle

%
%

\def\ts{t{\ss}s}
\def\ss{\hspace{.5pt}}
%
%
%
%

\def\i2{\mbox{\scriptsize\rm \"l\hspace*{-2.05pt}l}}

\def\bi2{\mbox{\footnotesize\rm \bf \"l\hspace*{-2.75pt}l}}

\def\ii2{\mbox{\footnotesize \it \"l\hspace*{-2.75pt}l}}

\def\I2{{\rm \"{I}}}

\newfont{\cyrfnt}{wncyr7 scaled 1400}

\newfont{\cyrB}{wncyb10 scaled 1400}

\newfont{\cyrftit}{wncyr8 scaled 985}


\begin{abstract}
Here we thoroughly discuss
some  weak points of the thermal model which is traditionally  used to  describe the hadron multiplicities measured in  the central nucleus-nucleus collisions. In  particularly, the role of  conservation laws, the values of  hard-core radii along with  the effects of the  Lorentz contraction of hadron eigen volumes   and the hadronic surface tension  are  systematically studied.
It is shown that for the adequate   description of  hadron multiplicities the   conservation laws should be modified, whereas  for the description of  hadron yield ratios  the conservation laws are not necessary at all. Also here we analyzed the usual criteria for the chemical freeze-out and found that none of them is robust.
A new chemical freeze-out  criterion of  constant entropy per hadron  equals to 7.18 is suggested and a novel effect of adiabatic chemical hadron production is discussed.
Additionally, we found that  the data
for the center of mass  energies above 10 GeV lead to  the  temperature of  the  nil hadronic surface tension coefficient of about   $T_0 = 147 \pm 7$ MeV. This  is a very intriguing result since a very close estimate for
such a temperature was obtained recently within entirely different approach.
We argue that
these two independently obtained results evidence that the (tri)critical temperature of the
QCD phase diagram is between 140 and 154 MeV.  In addition, here  we suggest to consider the pion and kaon hard-core radii as new fitting parameters. Such an approach for the first time  allows us to simultaneously describe  the hadron multiplicities and the Strangeness Horn and get a very high quality fit of the available experimental data.
\end{abstract}



\vspace*{-5mm}
\section{Introduction}
\vspace*{-2mm}

Experimental data on heavy ion collisions has traditionally been described by the thermal model [1-27]. The thermal model core assumption is that fireball produced in the relativistic nuclear collision reaches thermodynamic equilibrium. Such an assumption allows one to describe the multiplicities of  particles registered in the experiment using two parameters, namely the   temperature $T$ and baryo-chemical potential $\mu_b$.
The extracted values of $T$ and $\mu_b$ not only describe the experimental data, but they also give an essential information about the last stage of fireball evolution,  when  the inelastic collisions cease to exist, but the elastic collisions between hadron and the decay of resonances take place. This stage is usually called the chemical freeze-out.  The thermal model was initially used for the AGS
and SPS
data \cite{AGS_SPS} and was subsequently employed to describe the data collected at SIS  \cite{SIS1,SIS2}, SPS \cite{SPS} and  RHIC  \cite{RHIC1,RHIC2, RHIC3, RHIC4}. Using the thermal model it was possible to correctly predict   the hadron ratios measured  at LHC \cite{Andronic_big}, while  the only wrong prediction for LHC  was made for   $\bar p$/$\pi^-$ ratio \cite{LHC}. An analysis of the energy dependence of thermal parameters extracted from fits of the experimental data established the line of chemical freeze-out \cite{BM_Stachel_on_freezeout}. Consequently, thermal model is an established tool for particle production analysis and chemical freeze-out investigation.

\par However, the thermal model suffers from several weak points which should be accounted for in more careful studies. The present paper is just devoted to a critical analysis of the thermal model and contains several directions  to develop and  improve it.  First of all we would like to notice that the term "thermal model" is a common name for a set of similar models, each having its specific features such as the strangeness suppression factor $\gamma_s$, the inhomogeneous freeze-out scenario \cite{SIS1, RHIC3, Beccatini,  Dumitru}  etc.  Here we consider the minimal thermal model with two major parameters $T$ and $\mu_b$, following the approach of Andronic et al., \cite{Andronic_big}, and below we  briefly formulate the  main problems of the model to be analyzed in the present work.

\begin{itemize}
\item
\textbf{Particle table.}
In order to describe the experimental data using the thermal model one needs the masses, the widths and the decay branching ratios of all existing  resonances.
In principle, the mass spectrum of the hadronic resonances that are  heavier than 2,3
GeV is known poorly and, hence, they could create  a problem.
However, recently  it was shown that the large width of  heavy resonances
leads to their strong suppression \cite{Bugaev_why_we_see_no_Hagedorn_spectrum} and, hence, their contribution into
the thermodynamic functions of hadronic phase is negligible.

Note that not only the parameters of  hadrons from the "tail"  of mass spectrum are poorly known. For example, both the  mass  and the width of  $\sigma(600)$ meson are not well established, but the   thermal model predictions are strongly  influenced by
the values of  the  mass  and width of this meson
\cite{Strangeness_horn_sigma_meson}, while for many other baryons the width and branching ratios of decays are not well established at all. Thus, the particle table  is one source of uncertainty of the thermal model.

\item
\textbf{Hard-core spheres radii value.}
The ideal gas description has proven to be unsatisfactory \cite{Yen_Gorenstein_ideal_gas_vs_VdW} long ago. The simplest way to introduce an interaction between hadrons is to use the  repulsive  hard-core potential, since  the attraction between them is usually accounted for  via many sorts of hadrons \cite{Bugaev_why_we_see_no_Hagedorn_spectrum}. In the simplest case this  potential depends only on  a single parameter - the hard-core radius. In general case, each particle type may  have  its own hard-core radius. But for the sake of convenience and simplicity the hard-core radius is usually taken equal for all particles.  The usual  value for such a radius is $r=0.3$ fm. This value is motivated by the hard-core volume known from nucleon-nucleon scattering \cite{Thermal_model_review}. There are, however, two restrictions on the hard-core radii range: (i)  they should be small enough to satisfy the condition $V_{eigen}\ll V$, i.e. that the total  eigen volume of all particles $V_{eigen}$ should be much smaller than total volume of the system $V$; (ii)  on the other hand, these radii  shouldn't be too small, because otherwise the model will  lead to  a  contradiction with  the lattice quantum chromodynamics (QCD) thermodynamics data  \cite{Satarov:2011}.  Thus,   there is a  certain  freedom in defining the hard-core radii values which,  so far, was not systematically exploit  to describe the whole massive of existing experimental data.

\item
\textbf{Conservation laws.} Also here
we would like to discuss the baryon charge and isospin projection conservation laws. It was suggested to use them
in the form \cite{Andronic_big}
$$
\begin{cases}
 \sum_i n_i I_{3i}=I_{3 init}/V \,, \\
 \sum_i n_i B_i=B_{init}/V  \,.
\end{cases}
$$
The initial values are chosen $I_{3 init} = -20$ and $B_{init} = 200$  \cite{Andronic_big}, neglecting the fact that only the part of initial particles belong to the midrapidity region. Below we study the role of these conservation laws and  show that such a treatment leads to  physically unrealistic  freeze-out volumes  and to very  bad description of hadron  multiplicities, while the particle yield  ratios  (we use such a term  for the ratio of multiplicities in order avoid a confusion) description can be  extremely  good.

\item
\textbf{Multiplicities fit.}
The fit of  hadron multiplicities  is usually performed using  three parameters: $T,\mu_b,$ and $V$ \cite{Andronic_big}. Below we show that such a procedure combined with above mentioned conservation laws is mathematically ambiguous and it leads to the problems with the imposed baryonic charge conservation law.
\end{itemize}

In addition to these usual features of the thermal model we would like to thoroughly investigate the role of the Lorentz contraction of eigen
volumes of hadrons  to essentially improve the previous analysis \cite{Bugaev_Lorentz_cont_1, Bugaev_two_comp_VdW,Bugaev_Lorentz_cont_2} and to study the influence of the hadronic surface tension on
the fit of the freeze-out parameters.  As we argue the latter may provide us with new information about  the critical
temperature value of the QCD  phase diagram.  Such a comprehensive analysis of the different features of the thermal model and the
experimental data will also allow us to elucidate the correct  criterion of the chemical freeze-out which is a very hot topic nowadays.  Furthermore, we perform the simultaneous fit of hadron multiplicities and  the $K^+$/$\pi^+$ ratio for all available energies of collisions and, hence,   for the first time obtain the high quality fit of the
Strangeness Horn, i.e.   the peak in the $K^+$/$\pi^+$ ratio.

The work is organized as follows.
The basic features of the thermal model are outlined in the next section.
 In Section 3 we discuss the  important problems related to the conservation laws and propose their solutions. The discussion of the existing chemical freeze-out criteria and formulation of  a  new criterion
 along with a novel effect  of the adiabatic  chemical  hadron production is given in Section 4,
 while Section 5 is devoted to the model reformulation for the multi-component hadron gas mixture.
 In Section 6 we   investigate the values  of the hard-core radii and study the effect of the  Lorentz contraction of hard-core spheres.
 Section 7
is devoted to the analysis of the hadronic surface tension, while in Section 8 we describe the Strangeness Horn. Section 9 contains our conclusions.


\section{Model formulation}

In order to study the role of conservation laws in a form suggested in \cite{Andronic_big} and employed in their subsequent publications it is, first of all,  necessary to reproduce the results obtained in that work.
For this purpose
let us consider the Boltzmann gas consisting of $s$ sorts of hadrons having the temperature $T$ and the  volume $V$. Each $i$-th  sort  is characterized by its own  mass $m_i$ and chemical potential $\mu_i$. Suppose that the number of particles of i-th sort is $N_i$. Then its  canonical partition function is
\begin{eqnarray}\label{EqI}
&&Z_{can}(T,V,N_1,\dots,N_s) = \nonumber \\
&&= \prod\limits_{i=1}^s \left[ \frac{g_iV}{(2\pi)^3 }\int  \exp\left(-\frac{\sqrt{k^2+m_i^2}}{T} \right)d^3k\right]^{N_i}  \,.
\end{eqnarray}
Here $g_i = 2S+1$ is the degeneracy factor of i-th hadron sort, $k$ is the particle momentum. The corresponding grand canonical partition function reads as
\begin{eqnarray}\label{EqII}
Z_{gr.can.}&=&\sum_{N_1=0}^{\infty}\dots\sum_{N_h=0}^{\infty} \exp\left[\frac{\mu_1N_1+\dots+\mu_sN_s}{T} \right] \times  \nonumber \\
&\times &Z_{can}(T,N_1,\dots,N_s) \,.
\end{eqnarray}
From (\ref{EqII}) one gets the number of particles of each sort:
\begin{eqnarray}\label{EqIII}
N_i &=& V \, \phi_i(T,m_i,g_i) \, \exp \left[\frac{\mu_i}{T}\right] \,  \nonumber \\
&\equiv & \frac{g_iV}{(2\pi)^3}\int \exp\left(\frac{\mu_i -\sqrt{k^2+m_i^2}}{T} \right)d^3k  \,.
\end{eqnarray}
Following the commonly accepted approach, we consider the  conservation of baryon charge $B$, strangeness $S$ and isospin projection $I_3$ on average:
\begin{eqnarray}\label{EqIV}
\sum_{i=1}^N n_i S_i = S_{init} = 0\,,\\
\label{EqV}
\sum_{i=1}^N n_i B_i  = B_{init}/V = 200/V\,,\\
\sum_{i=1}^N n_i I_{3i} = I_{3init}/V = - 20/V \,.
\label{EqVI}
\end{eqnarray}
These conservation laws define the value of the total chemical potential for
the hadron of sort $i$ as $\mu_i=B_i\cdot \mu_b + S_i\cdot \mu_s + I_{3i}\cdot \mu_{I_3}$, where the quantities  $B_i$, $S_i$ and $I_{3i}$ denote, respectively,
the baryonic, strange and isospin projection of  such a hadron, while the corresponding chemical potentials are denoted as $ \mu_b$, $ \mu_s$ and $\mu_{I_3}$.

The interaction of hadrons and resonances is usually accounted for  by the  hard-core repulsion of  the Van der Waals type  \cite{Yen_Gorenstein_excluded_volume} as
\begin{eqnarray}\label{EqVII}
p=p_{id. gas}\cdot \exp\left(- \frac{p\cdot b}{T} \right), \quad
n_i  = \frac{ n^{id}_i \exp\left(- \frac{pb}{T} \right)}{1+\frac{pb}{T}} \,,
\end{eqnarray}
where the pressure $p_{id. gas}$ and the $i$-th charge density $n^{id}_i$  of
an ideal gas is modified  due to hard-core repulsion.
Here $b=\frac{2 \pi}{3}(2R)^3$ is the excluded volume for  the hard-core  radius $R$, which in actual calculations was taken to be $R=0.3$ fm for all hadrons.
The usual Van der Waals correction affects  the particle densities, but has no effect on particle ratios \cite{Yen_Gorenstein_excluded_volume}. While  its
effect on the charge and particle  densities may be strong, for the freeze-out
densities  obtained at  and above the highest  SPS energy $\sqrt{S_{NN}} = 17.6$  GeV
the excluded volume correction leads to a reduction of the densities by about  of  50 percent.

\par The resonance decays are usually  accounted for in the following way: the final
 multiplicity of hadron $X$  consists of the thermal contribution $N^{th}_X$  and the decay ones:
\begin{eqnarray}\label{EqVIII}
N^{fin}_X = N^{th}_X+ N^{decay} = N^{th}_X + \sum_{Y} N^{th}_Y \, Br(Y \to X) \,,
\end{eqnarray}
where $Br(Y \to X)$ is the decay branching of  the Y-th hadron  into the hadron X. The masses, the  widths and the decay branchings were  taken from the particle tables  used  by  the  thermodynamic code THERMUS \cite{THERMUS}.

\par The width $\Gamma$ of the resonance of mean  mass $m$  is   accounted for  by replacing the Boltzmann
distribution function in the particle  pressure  by its average over the
 Breit-Wigner mass distribution as
\begin{eqnarray}\label{EqIX}
&&\int \exp\left(\frac{-\sqrt{k^2+m^2}}{T} \right)d^3k \rightarrow \nonumber \\
&& \rightarrow \frac{\int_{M_0}^{\infty} \frac{dx_i}{(x_i-m)^2+\Gamma^2/4}\int \exp\left(\frac{-\sqrt{k^2+x_i^2}}{T} \right)d^3k}{\int_{M_0}^{\infty} \frac{dx_i}{(x_i-m)^2+\Gamma^2/4}} \,,
\end{eqnarray}
where $M_0$ is the  dominant decay channel mass.

\section{Role of  conservation laws}

\par Using the thermal model  formulated in previous section   we fitted the hadron yield  ratios in the  energy range from AGS to RHIC, i.e. for  $\sqrt{S_{NN}}$ = 2.7 $\div$ 200 GeV.    We used  the $\chi^2$ minimization for all the ratios available for this energy range as the fit criterion.  The present consideration is very similar to that one used in \cite{Andronic_big}.  The main sources of difference are listed below.
\begin{itemize}
\item
   \textbf{The Boltzmann statistics} is used here  instead of the quantum statistics
   employed  in  \cite{Andronic_big}. This allows us to essentially fasten the simulations since the  momentum integration can be done  only once for each  hadron species. We checked that for the freeze-out temperatures  $T\ge 50 $ MeV
   obtained here the difference of the results due to the Boltzmann  statistics  is almost negligible.
\item

 \textbf{The charm conservation} is not accounted for by the present model, since  it is important only for the charmed particles multiplicities description which is not considered here.
\item
    \textbf{The particle table} used here is slightly different from that one of \cite{Andronic_big}, but this does not lead to a big difference in results.   Although,  in contrast to
    \cite{Andronic_big},   we do not fit the mass and the width of $\sigma(600)$ meson.
\item
    \textbf{Inclusion of  the resonance width} is done in this work  for all values of colliding energy  $\sqrt{S_{NN}}$, while in \cite{Andronic_big} the width was accounted only  for  the  AGS energy range.
\end{itemize}


\begin{figure}[htbp]
  \centerline{
    \includegraphics[height=6cm]{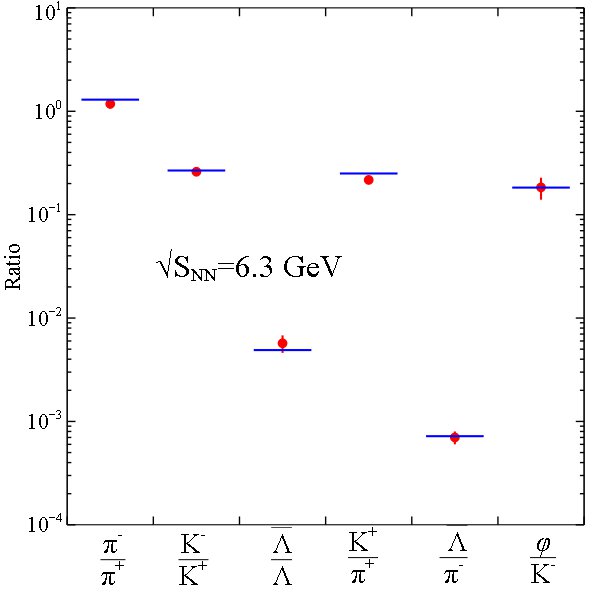}
  }
  \centerline{
   \includegraphics[height=6 cm]{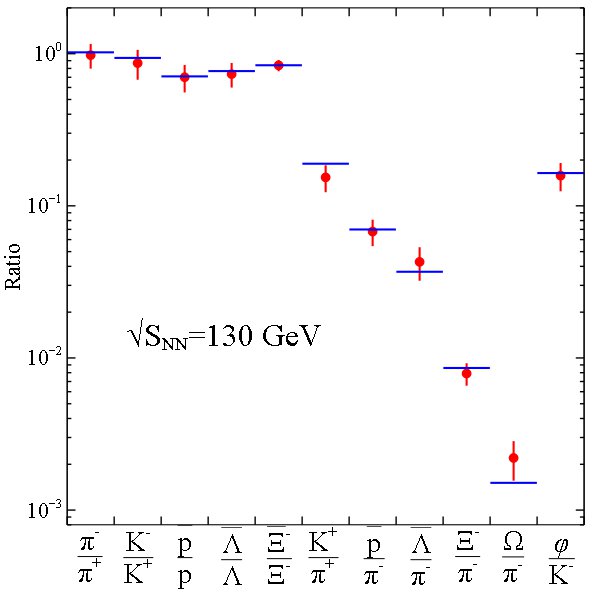}
  }
 \caption{The examples of the particle  yield ratios description. The dots denotes the experimental values,  while the lines show the result of fit. Upper panel:  $\sqrt{S_{NN}}$ = 6.3 GeV, T=139 MeV, $\mu_b$ = 503 MeV, the mean square deviation per degree of freedom is $\chi^2/NDF = 4.8/4$.
Lower panel: $\sqrt{S_{NN}}$ = 130 GeV, T=169 MeV, $\mu_b$ = 31 MeV, $\chi^2/NDF = 3.4/9$.}
  \label{ratios_with_width}
\end{figure}

\begin{figure}[htbp]
    \centerline{
    \includegraphics[height=6cm]{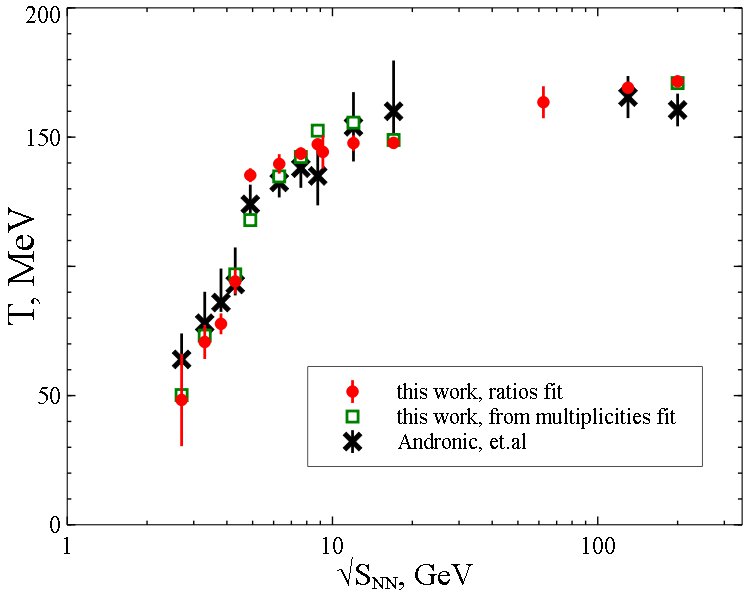}
  }
    \centerline{
   \includegraphics[height=6 cm]{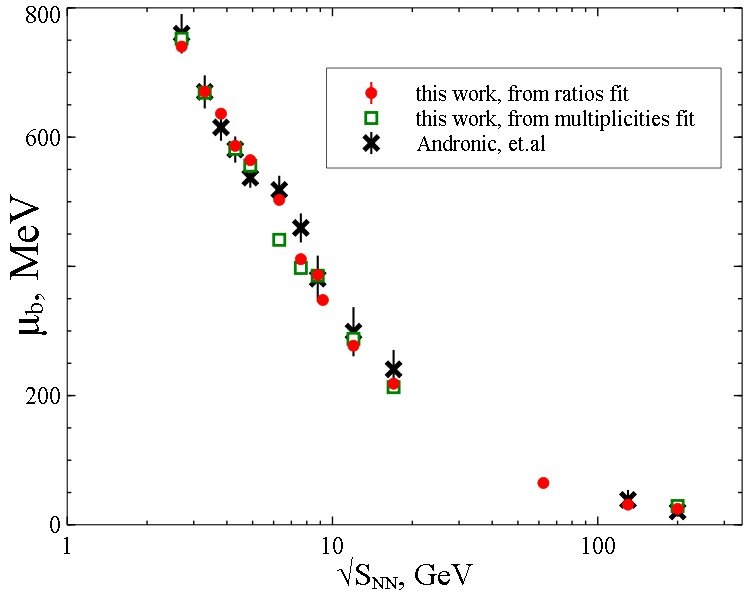}
  }
 \caption{Dependence of thermal model fitting parameters on the center of mass collision energy $\sqrt{S_{NN}}$. Upper panel:  the chemical freeze-out temperature $T$ vs. $\sqrt{S_{NN}}$.
Lower panel: the chemical freeze-out baryonic chemical potential  $\mu_b$ vs. $\sqrt{S_{NN}}$.  The   results obtained from  the fit of hadron yield ratios (circles) with the conservation laws
and from the fit of hadron multiplicities (open squares)
are compared with that ones obtained in \cite{Andronic_big} (crosses).}
  \label{Thermal_par_Andr_like}
\end{figure}

As it is seen from Fig. \ref{ratios_with_width} the experimental hadron yield  ratios are reproduced very well within the present model.
The dependence of   the chemical freeze-out fitting  parameters on $\sqrt{S_{NN}}$ is given  in Fig.  \ref{Thermal_par_Andr_like}.
In Fig.  \ref{Thermal_par_Andr_like} we  show
only  the statistical errors for the obtained fit, while for the  parameters of work \cite{Andronic_big}
the shown errors account for  the systematic and the statistical ones. As one can see from
Fig.  \ref{Thermal_par_Andr_like} the discrepancy between the results of the present model and that one of  \cite{Andronic_big} is within the error bars.

\par
\begin{figure}
  \centerline{
\includegraphics[height=6cm]{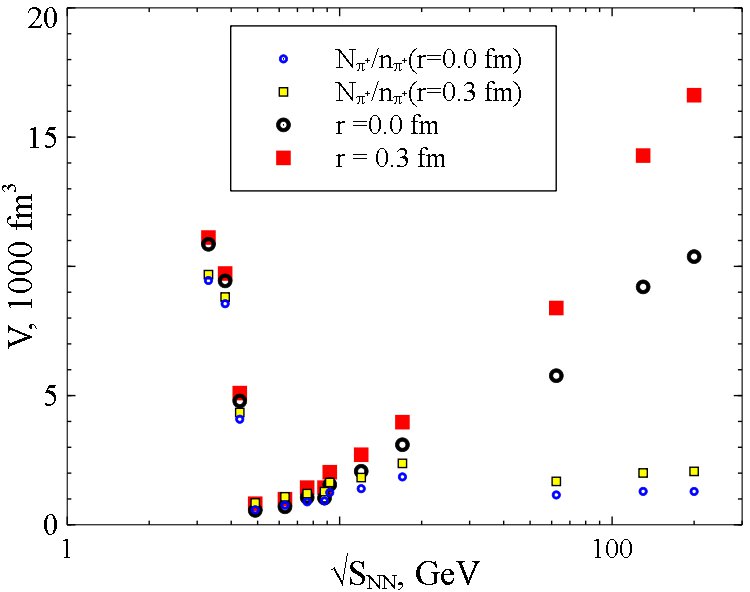}}
 \caption{The chemical freeze-out volume vs. $\sqrt{S_{NN}}$ for the ideal hadron gas and the hadron gas with the  hard-core radii 0.3 fm. The smaller symbols correspond to the fit of hadron yield ratios with all the conservation laws (\ref{EqIV})-(\ref{EqVI}) accounted for, while the larger symbols are obtained by the fit of hadron multiplicities ignoring Eq. (\ref{EqV}) (see text for details).} \label{Volume_huge}
\end{figure}

 Similarly to \cite{Andronic_big} we found that  both the chemical freeze-out  temperature $T$ and  baryonic chemical potential $\mu_b$ are almost independent of  the initial value of the  baryon charge $B_{init}$ and the initial value of the isospin projection $I_{3_{init}}$. However, we found that the freeze-out volume $V$ which stands on the right part side of conservation laws (\ref{EqV}) and (\ref{EqVI})
 is very sensitive to them. The obtained  chemical freeze-out  volume dependence on $\sqrt{S_{NN}}$ for $B_{init} = 200$ and $I_{3_{init}} = -20$ is shown in  Fig.  \ref{Volume_huge}.

From the larger symbols in Fig.  \ref{Volume_huge}  one can clearly see that
for the center of mass collision energies  $\sqrt{S_{NN}} = 2.7 - 4.3$ and
$\sqrt{S_{NN}}= 12 - 200$ GeV the found chemical freeze-out volume is so large  that
it  exceeds the volume of  kinetic freeze-out \cite{Kinetic_freezeout}.
Here we found that
unlike the hadron yield  ratios, the chemical freeze-out volume is very sensitive both to the excluded volume correction and to the values of  parameters $B_{init}$ and $I_{3_{init}}$. From Eq. (\ref{EqV}) one can deduce that the larger value of the
excluded volume $b$ corresponds to  the larger value of the chemical freeze-out volume $V$ since larger $b$ value  one obtains the smaller particle concentrations $n_i$ and, consequently, the  larger volume $V = B_{init}/\sum{n_iB_i}$. Therefore,  the minimal chemical freeze-out  volume corresponds to an ideal gas, i.e. for  $b=0$. This minimal chemical freeze-out  volume is also  shown in Fig.  \ref{Volume_huge}. Despite the absence of the excluded  volume correction, chemical freeze-out  volume for an ideal gas  remains huge.  Hence, we conclude that the initial values in conservation laws, namely $B_{init}$ and $I_{3_{init}}$ are of crucial importance  for an extraction of  chemical freeze-out volume.

\par
\begin{figure}[htbp]
  \centerline{
    \includegraphics[height=6cm]{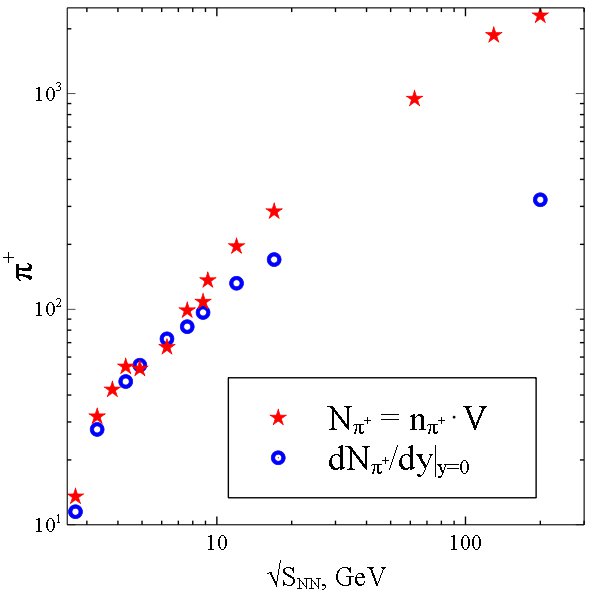}
  }
    \centerline{
   \includegraphics[height=6 cm]{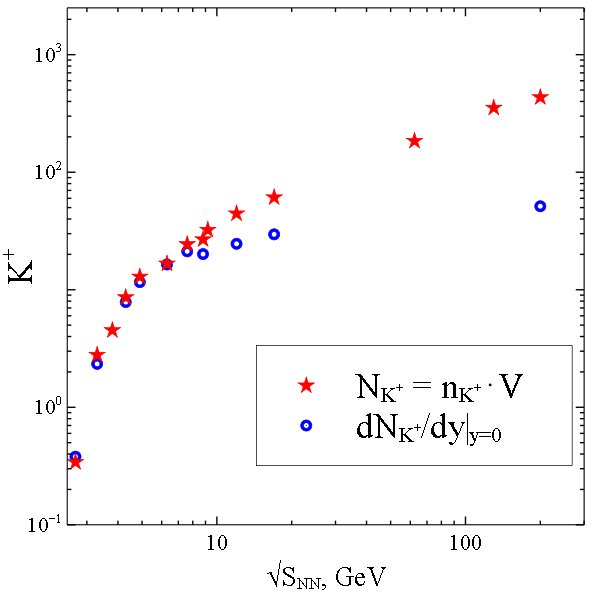}
  }
 \caption{Upper panel:  $\pi^+$ multiplicity at chemical freeze-out vs. $\sqrt{S_{NN}}$.
Lower panel: $K^+$ multiplicity at chemical freeze-out vs. $\sqrt{S_{NN}}$. The circles correspond to  the experimental data, whereas the stars are found from particle densities as $N = n \cdot V$}
  \label{Mult_huge}
\end{figure}

  From the comparison of  the  hadron multiplicity $N = n \cdot V$ obtained by the thermal model
 and  its experimental value $dN/dy|_{y=0}$ measured at zero rapidity,  one can conclude whether  the thermal model provides a reliable description  of hadron multiplicities.  Such a comparison for $\pi^+$ and $K^+$ mesons is shown  in Fig. \ref{Mult_huge}. Since the hadron yield ratios shown in Fig. 1  are described well by the thermal model, then one would expect that , if the multiplicity of a single  hadron type   is in a good agreement with the  experiment data, then the multiplicities of all other sorts
 of hadrons  should be well  described  too.
 However, from Fig.  \ref{Mult_huge} one can see that at $\sqrt{S_{NN}} \ge$ 10 GeV the experimental values of multiplicities are much smaller than the theoretical ones. At lower $\sqrt{S_{NN}}$ there is no such a problem, despite the big chemical freeze-out   volume
 $V$. Our conclusion  is that for higher energies  the value of parameters $B_{init}$ and $|I_{3_{init}}|$ should be taken smaller than for lower energies. In particular, according to Fig.  \ref{Mult_huge} at $\sqrt{S_{NN}}$ = 200 GeV the value of $B_{init}$ and $|I_{3_{init}}|$ should be  about 10 times smaller than tat ones for the low collision energies.

\begin{figure}
  \centerline{
\includegraphics[height=6cm]{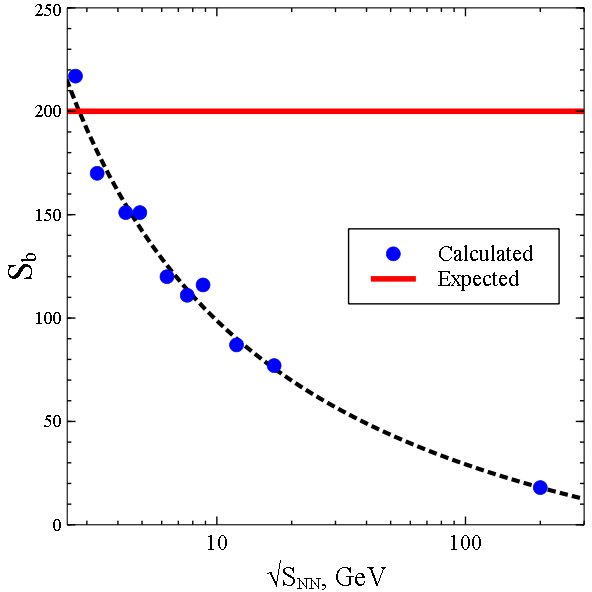}}
 \caption{The obtained baryonic charge $S_b = V \cdot \sum_{i=1}^N n_i B_i$ (dots)  vs. $\sqrt{S_{NN}}$. The dots  are  calculated from the baryon charge conservation, whereas the  line corresponds to the expected value $B_{init} = 200$.} \label{bar_sum}
\end{figure}

A different way to describe the hadron multiplicities was used in \cite{Andronic_big}. The chemical freeze-out volume $V$ was treated there as a free parameter. Let's show that such a treatment leads to mathematical ambiguity. Consider the conservation laws (\ref{EqIV})--(\ref{EqVI}) together with
the following expression for system pressure
\begin{eqnarray}\label{EqX}
p = p^{id}(T, \mu_b, \mu_s, \mu_{I_3})\cdot  \exp\left(- \frac{p\, b}{T} \right)\,.
\end{eqnarray}
Such a system of equations  has six unknowns, i.e.  $T$, $\mu_b$, $\mu_s$, $\mu_{I_3}$, $V$, $p$,  and four  equations. Hence, two unknowns should be treated as free fitting parameters. If , however, one  treats three unknowns as free parameters,
then one of the equations may be not satisfied in general.
More specifically, if $T$, $\mu_b$ and $V$ are the  free fitting parameters,  then the baryon charge conservation equation (\ref{EqV}) may be broken down.
To demonstrate this explicitly we have considered the thermal model fit  with three free parameters - $T$, $\mu_b$ and $V$ and ignored  the  baryon charge conservation equation (\ref{EqV}),
while the isospin projection conservation law  (\ref{EqVI}) was used to find the chemical potential $\mu_{I_3}$.
After  fitting the experimental hadron multiplicities  $dN/dy|_{y=0}$ (not their ratios!) for the same energy range as before, we  found the  resulting baryonic charge as  $S_b = V \cdot \sum_{i=1}^N n_i B_i$ summing up   the  densities  $n_i$ of all  baryons  and anti-baryons  multiplied by their  baryonic charge $B_i$.
Clearly, if  Eq. (\ref{EqV})  is satisfied, then this sum $S_b $ should match the value of $B_{init} = 200$. Fig. \ref{bar_sum}, however,  demonstrates that Eq. (\ref{EqV})  cannot be satisfied.
Although the chemical freeze-out temperature and baryonic chemical potential obtained by the fitting
of the hadron multiplicities do not differ essentially from that ones found  by the fit of hadron yield
ratios (see open squares in Fig. \ref{Thermal_par_Andr_like}), the   freeze-out volumes obtained  from the multiplicity fit  are essentially smaller (see smaller symbols in Fig. \ref{Volume_huge}) and more physically adequate for $\sqrt{S_{NN}}\ge 5$ GeV  than that ones found from the fit of
hadron yield ratios.

Additionally, here we found that the values of  $B_{init}$ and $I_{3_{init}}$
should strongly  depend on collision energy: at $\sqrt{S_{NN}}$ = 200 GeV they are about ten times smaller than at the AGS energies.
The hadron yield ratios are not sensitive to the values of $B_{init}$ and $I_{3_{init}}$, and hence   the baryon charge  and $I_3$ conservation  can be neglected
for
the description of hadron yield  ratios. If, however, one supposes that $B_{init}=const >0$ and $I_{3_{init}}=const >0$, then the description of hadron multiplicities completely fails.
Evidently, one can describe the hadron multiplicities by  introducing $\sqrt{S_{NN}}$ dependence of  $B_{init}$ and $I_{3_{init}}$ values,  if such  dependences  are  known. Since such dependences are unknown, then one has to
ignore the baryon charge (\ref{EqV}) and isospin projection (\ref{EqVI}) conservation laws,  and to fit the parameters $T$, $\mu_b$, $\mu_{I_3}$ and $V$ to describe the hadron multiplicities or to fit the parameters $T$, $\mu_b$, $\mu_{I_3}$ in order to get the  description of hadron yield ratios. Note, however, that the
strangeness conservation laws (\ref{EqIV}) does not create such problems and, hence, it should be always obeyed.


\section{Chemical freeze-out criteria and adiabatic chemical  hadron production}

\begin{figure}[htbp]
  \centerline{
    \includegraphics[height=6cm]{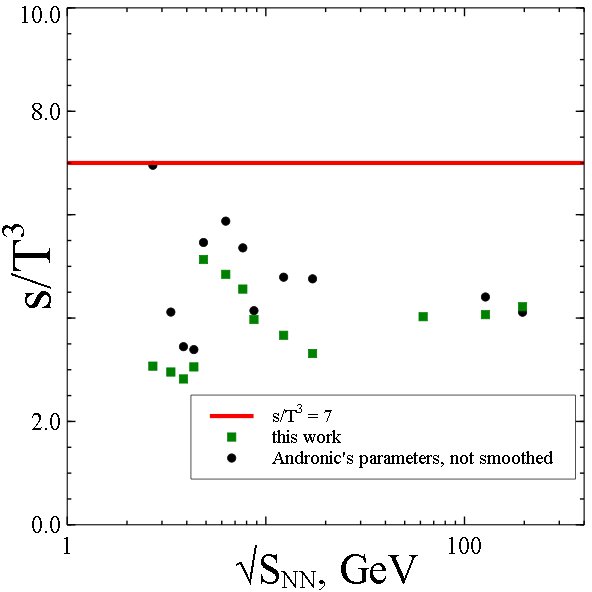}
  }
    \centerline{
   \includegraphics[height=6 cm]{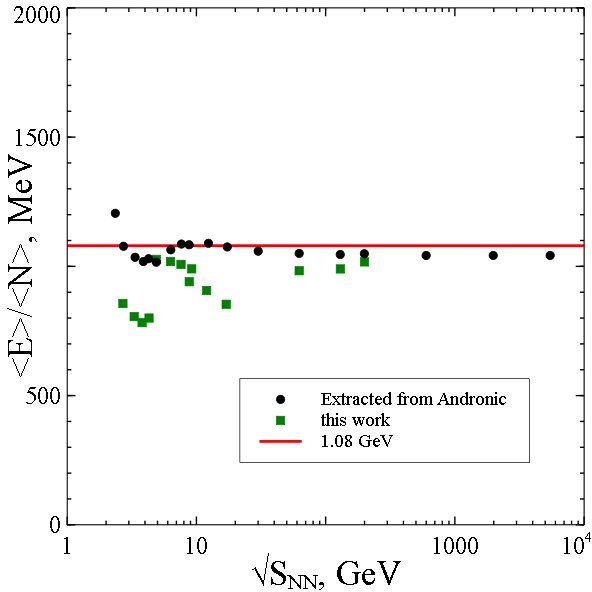}
  }
      \centerline{
   \includegraphics[height=6.2 cm]{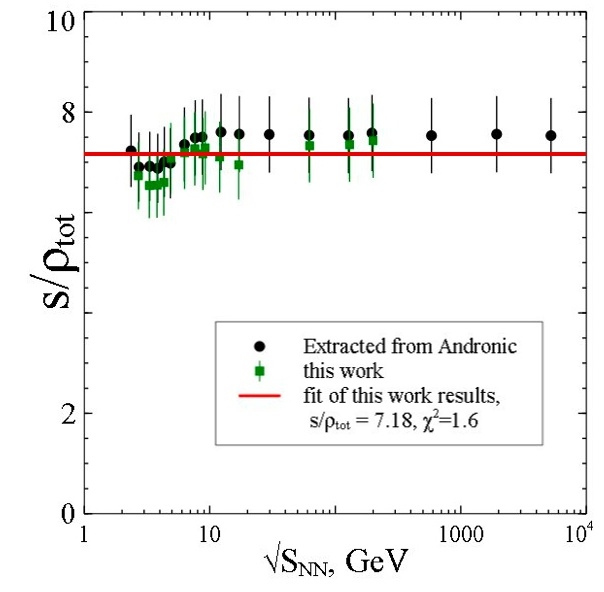}
  }
 \caption{Different chemical freeze-out criteria. Upper panel: ratio of the entropy density to the cube of temperature $s/T^3$ at chemical freeze-out vs. $\sqrt{S_{NN}}$.
Middle  panel: energy per particle  $\langle E \rangle / \langle N \rangle$  at chemical chemical freeze-out vs. $\sqrt{S_{NN}}$.
Lower panel: the novel criterion of chemical freeze-out, entropy per particle at chemical freeze-out $s/\rho \simeq 7.18$. The shown errors  are combined the statistical and systematic errors.
The results of present work (squares) are  very similar with that ones extracted from \cite{Andronic_big}.}
  \label{FO_crit}
\end{figure}

The thermal model discussed above allows us to clarify the long standing question on the physically
appropriate chemical freeze-out criterion which is widely discussed \cite{BM_Stachel_on_freezeout,
ThermalFO}. The most popular chemical freeze-out criteria are (I)  the constant value of the mean energy per hadron, $\langle E \rangle / \langle N \rangle \simeq 1.08$ GeV, (II)  the constant value of the
entropy density to the cube of the temperature, $s/T^3 \simeq 7$,  and (III) the constant value of  a total baryon and antibaryon density $n_B + n_{\bar B}\simeq 0.12  $  fm$^{-3}$.
The criterion (I)  is believed to be more robust, while  the criteria (II) and (III) show strong
dependence on the hard-core radius value \cite{ThermalFO}. We have performed the analysis and found the criteria (II) and (III) are  not obeyed at all, while the  criterion (I) validity
depends essentially  on the thermal model parameterization.    The validity of these statements for  the criteria (I) and  (II) are, respectively, demonstrated in the middle and the
 upper panels of Fig.  \ref {FO_crit}. Moreover, the thermal model  results that we extracted from
 \cite{Andronic_big} are very similar to our ones, despite several  differences in the parameterization of   these two models.

Our detailed analysis shows that there exists a much more robust chemical freeze-out  criterion than all previously discussed ones. This novel  criterion corresponds to  the constant value of the entropy per number of particles which  in terms of the entropy density $s$  and hadron number  density $\rho_{part}$ can be expressed as
\begin{equation}\label{EqNew}
\frac{s}{\rho_{part}} \simeq 7.18 \,.
\end{equation}
The lower panel of Fig. \ref{FO_crit} shows that for two different parameterizations of the thermal model the ratio  $\frac{s}{\rho_{part}}$ stands between 6.6 and 7.6, i.e. the deviates  within  $\pm~  8$~\% only,   while the values of the center of mass energy of collision change for two orders of magnitude!
Such a behavior of  the $\frac{s}{\rho_{part}}$ quantity  evidences for the {\bf adiabatic chemical  hadron production} in heavy ion collisions.

\section{Multi-component gas and hard-core radii}

Although  it is clear that the hadron radii can serve as the  parameters of the thermal model, they, however,  are rarely treated as the free parameters. The common approach is to fix one radius for all hadrons. The  value of this radius was discussed \cite{Heppe_radii} and it ranges  from 0.2 fm to 0.8 fm \cite{big_radii}.
However, the value  $r$=0.3 fm that was taken from nucleon-nucleon scattering \cite{Nucl_nucl_scattering} seems to be an established value.
To investigate the role of the hard-core  radii we introduce  the different radii for mesons and for baryons $R_m$ and $R_b$, respectively. Also to  study an influence of the Lorentz contraction of hard-core radii  on the chemical freeze-out parameters  $T$, $\mu_b$ and on the hadron yield particle ratios we need the hadron gas  model
which accounts for  different values of their  eigen volumes.
 For this purpose we  use  an approach developed in \cite{Bugaev_Lorentz_cont_1,Bugaev_two_comp_VdW,
Bugaev_Lorentz_cont_2}.
Below we give the necessary theoretical apparatus to study  the  multi-component hadron gas mixture, whereas   the results of the global fit for the cases with and without Lorentz contraction are given in the subsequent section.
%
%
%
%
%
%
%

\par Consider again the Boltzmann gas of $s$ hadron species in a volume $V$ at a temperature $T$. Let $N_i$ be  a quantity of the i-th sort of hadrons
\begin{eqnarray}\label{EqXI}
N=
\begin{pmatrix}
 N_1 \\
 N_2 \\
... \\
N_s
\end{pmatrix}\,.
\end{eqnarray}
The total number of particles is $M=\sum_{i=1}^s N_i$. It is  assumed that for  every two sorts of hadrons  $i$ and $j$ there is  the  excluded volume $b_{ij}$. Then one can introduce the excluded volume matrix $B=(b_{ij})$. Naturally, it is supposed that the matrix $B$ is symmetric, i.e. $b_{ij}=b_{ji}$.

\par
The canonical partition function can be  obtained by  adding the particles of some eigen volume  one-by-one and taking in account all the  corresponding excluded volumes of the previously added particles.  Such an approximation  was suggested  in \cite{Bugaev_two_comp_VdW} and it gives the following expression for the
canonical partition of the Van der Waals hadron gas mixture
\begin{eqnarray}\label{EqXII}
Z_{VdW}(T,V,N_i)=\left[\prod_{i=1}^s \frac {\phi_i^{N_i}}{N_i!}\right]\times \left[V -\frac{N^TBN}{M} \right]^M
\end{eqnarray}
where the thermal particle density  $\phi_i(T,m,g)$ is defined in (\ref{EqIII}),
and $N^T$ is the transposed matrix to that one given by  (\ref{EqXI}).


In the next step we write the grand canonical partition function (GCPF) as
\begin{eqnarray}\label{EqXIII}
\mathcal{Z}=\sum_{N_1=1}^{\infty}\sum_{N_2=1}^{\infty}\dots\sum_{N_s=1}^{\infty}\left(\prod_{i=1}^s \exp \left[\frac{\mu_i N_i}{T} \right]\right) \times Z_{VdW} \,.
\end{eqnarray}
It is well known  \cite{Pathria} that in the thermodynamic limit the GCPF can be replaced by the maximal term of the multiple sum in  $\mathcal{Z}$ (the maximum term method). Suppose that the array $N^*$ gives the maximal term of $\mathcal{Z}$. Then the system pressure is given by
\begin{eqnarray}\label{EqXIV}
&& p/T=\lim_{V \to \infty}\frac{\mathcal{Z}}{V}=  \nonumber \\
&&\lim_{V \to \infty}\frac{1}{V}\ln\left[ \prod_{i=1}^s \frac {A_i^{N_i^*}}{N_i^*!}\times  \left (V-\frac{(N^*)^TBN^*}{ M^*} \right)^{ M^*}\right]\,,
\end{eqnarray}
where $A_i=\phi_i \exp \left[ \frac{\mu_i}{T} \right]$.
Let us find $N^*$ from the maximum conditions  ($i=1..s$):
\begin{eqnarray}\label{EqXV}
{\partial \over \partial N_i^*} \left[ \ln\left[ \prod_{i=1}^s \frac {A_i^{N_i^*}}{N_i^*!}  \left (V-\frac{(N^*)^TBN^*}{ M^*} \right)^{  M^*}\right]  \right]=0 . \,
\end{eqnarray}
Performing the differentiations, one gets
\begin{eqnarray}\label{EqXVI}
\xi_i=A_i \exp\left(-\sum_{j=1}^s 2\xi_j b_{ij}+\frac{\xi^TB\xi}{\sum_{j=1}^s\xi_j}\right) \,,
\end{eqnarray}
with $\xi_i=\frac{N_i}{V-\frac{N^TBN}{M}}$ and
\begin{eqnarray}\label{EqXVII}
\xi=
\begin{pmatrix}
 \xi_1 \\
 \xi_2 \\
... \\
\xi_s
\end{pmatrix}\,.
\end{eqnarray}
Using  (\ref{EqXVI}) one can express the hadron densities  $n_i = \frac{N_i^*}{V}$ and the system pressure $p$ as
\begin{eqnarray}\label{EqXVIII}
n_i=\frac{\xi_i}{1+\frac{\xi^TB\xi}{\sum_{j=1}^s \xi_j}},   \quad p = T \, \sum_{i=1}^s \xi_i
\end{eqnarray}
The solution of the system of equations (\ref{EqXVI})--(\ref{EqXVIII}) defines the
hadron densities   for the multi-component gas.

\par In a special case  when all  the elements of the excluded volumes matrix are equal  $b_{ij}=v_0$  Eqs. (\ref{EqXVI})--(\ref{EqXVIII})  give
\begin{eqnarray}\label{EqXIX}
\begin{cases}
\xi_i=A_i \exp(-p\,v_0/T) \,, \\
n_i=\frac{\xi_i}{1+p\,v_0/T} \,, \\
p/T= \left( \sum_{i=1}^s A_i \right) \cdot  \exp(-p\,v_0/T) \,.
\end{cases}
\end{eqnarray}
In this case the ratios of  two particle densities from (\ref{EqXIX})
match that ones of the mixture of the corresponding  ideal gases  for an arbitrary
value of $v_0$, while the particle densities themselves may essentially differ
from the particle densities of the ideal gas.

\section{Results for hard-core  radii  with the  Lorentz contraction}

In this subsection it is assumed that all baryons have the same hard-core radii $R_b$ and all mesons have the same  hard-core radii $R_m$. Then performing the global fit we would like, first,  to find the pair of radii $(R_m,R_b)$ that provides the best fit, and, second, we would like to study the influence of Lorentz contraction of the
chemical freeze-out parameters.

To simplify the numerics let us define that two hadrons  belong to the same  type, if their excluded  volumes are equal. The number of equations in the system (\ref{EqXVI}) is equal to the number of particle types. Hence, the case with the Lorentz contraction included  is more complicated, because instead of  two sets of particles one should treat each hadron type as a new kind of particles. In order to  avoid the large  number of  equations in system (\ref{EqXVI}), the  particles heavier than 900 MeV are considered  non-relativistically, i.e. they all belong to two sorts: either to  the baryons with the hard-core radius $R_b$ or to  heavy  mesons with hard-core radius $R_m$. For two particles $i$ and $j$, both heavier than 900 MeV, the element of excluded volume matrix $b_{i,j}$ is $\frac{2\pi}{3} (R_i+R_j)^3$. For other cases the second virial coefficients $b_{i,j}$ are  calculated using Eqs. (3) and (4) of \cite{Bugaev_Lorentz_cont_2}  by
the direct  translation of  one ellipsoid around the other  with the subsequent   averaging of the obtained excluded volume over the ellipsoid positions.


The fit procedure is the same as described in  the preceding subsection:
we fit the hadron yield ratios by $T$, $\mu_b$ and $V$, respect the isospin projection conservation law
(\ref{EqVI}) to find the value of  $\mu_{I_3}$, but  ignore
the baryonic charge  conservation law (\ref{EqV}).
This our effort  to study the role of  the Lorentz contraction is inspired by the fact that the conventional thermal model has problems with the light mesons description. Thus, to describe the pion multiplicity and the ratios containing  the pions it was proposed to introduce $R_{\pi}$ which is smaller  than all the other hadron radii \cite{R_pi1, Yen_Gorenstein_excluded_volume}. Another example of difficulties with the light mesons is a "Strangeness Horn", i.e. the peak in the $K^+$/$\pi^+$ ratio. Its description  was finally improved by fitting the  $\sigma(600)$ meson mass and width \cite{Strangeness_horn_sigma_meson}, but the obtained  description is far from being very good.

\par
Let us see whether  the Lorentz contraction might resolve these problems. At a given temperature the  eigen volume of the lighter particles decrease more than that one  of the heavier ones. Consequently, the excluded volume of  lighter particles gets smaller  compared to  the excluded volume of   heavier ones  \cite{Bugaev_Lorentz_cont_1,Bugaev_two_comp_VdW,
Bugaev_Lorentz_cont_2}. Such a behavior of the Lorentz contracted excluded volume   provides us with the natural explanation of the fact that pion hard-core radius $R_{\pi}$ is  smaller than other hard-core radii. On the other hand, this might  improve the Strangeness Horn description. It was also shown \cite{Bugaev_Lorentz_cont_2} that the Lorentz contraction removes  the causality paradox  from  the thermal  model, i.e. at high densities the speed of sound does not exceed  the speed of light.
Therefore, it is necessary to incorporate the Lorentz contraction into conventional thermal model and study its effect on the hadron multiplicity description. In order to make the numerical evaluation of the relativistic excluded volumes faster we heuristically  derived an approximative formula for such volumes  which allows one to reduce
the six dimensional integration over  the pair of  three-vectors of particle momenta to the three dimensional
integral. The derivation of such a formula and  its verification are  given in the Appendix.

\begin{figure}[htbp]
  \centerline{
    \includegraphics[height=7cm]{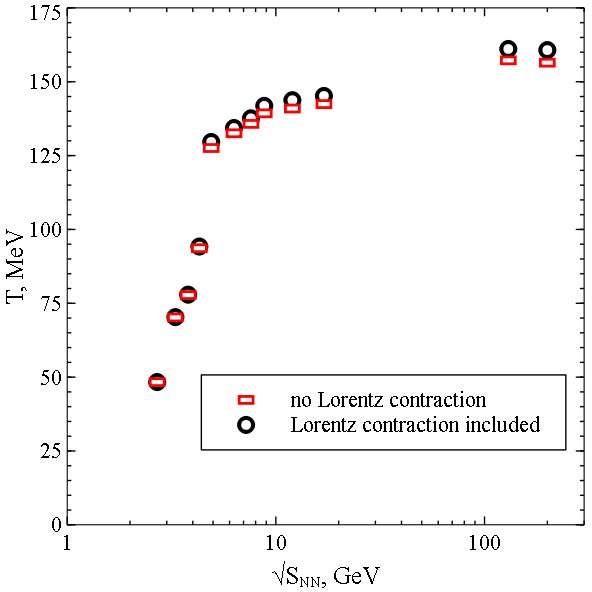}
  }
    \centerline{
   \includegraphics[height=7 cm]{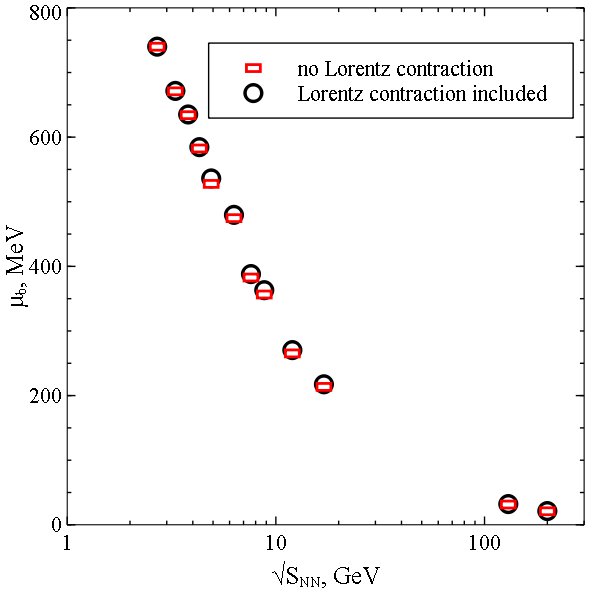}
  }
 \caption{The chemical freeze-out parameters obtained from the fit with (circles) and without (rectangles) the Lorentz contraction. Upper panel:  the chemical freeze-out temperature $T$ vs.  $\sqrt{S_{NN}}$.
Lower panel: the chemical freeze-out  baryonic chemical potential  $\mu_b$ vs.  $\sqrt{S_{NN}}$.}
  \label{T_mu_rel_nonrel}
\end{figure}

\begin{figure}
  \centerline{
\includegraphics[height=6cm]{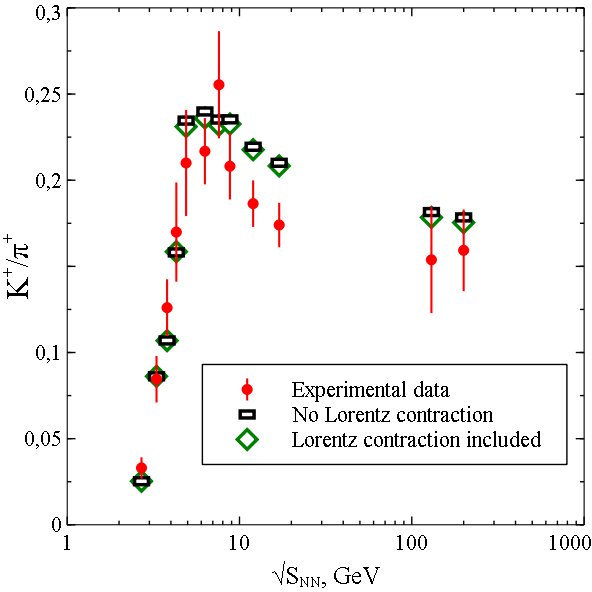}}
 \caption{Strangeness Horn description improvement for the model with the Lorentz contraction included and without it.}
 \label{str_horn}
\end{figure}

\par To compare  the models with and without the Lorentz contraction, we have chosen hard-core radii $R_m$=0.45 fm, $R_b$ = 0.3 fm and have found the new best-fit $T$ and $\mu_b$ values  for the case with Lorentz contraction, see Fig. \ref{T_mu_rel_nonrel}. From this figure one can conclude that the baryo-chemichal potential is almost unaffected by the Lorentz contraction,  while the temperature is slightly higher  for the case with the Lorentz contraction.
It is also  interesting to check, whether the inclusion of the Lorentz contraction improves  the Strangeness Horn description. From Fig. \ref{str_horn} we conclude that there is a small improvement, which  is not sufficient  to qualitatively improve the Strangeness Horn description.

\par
\begin{figure}
  \centerline{
\includegraphics[height=6cm]{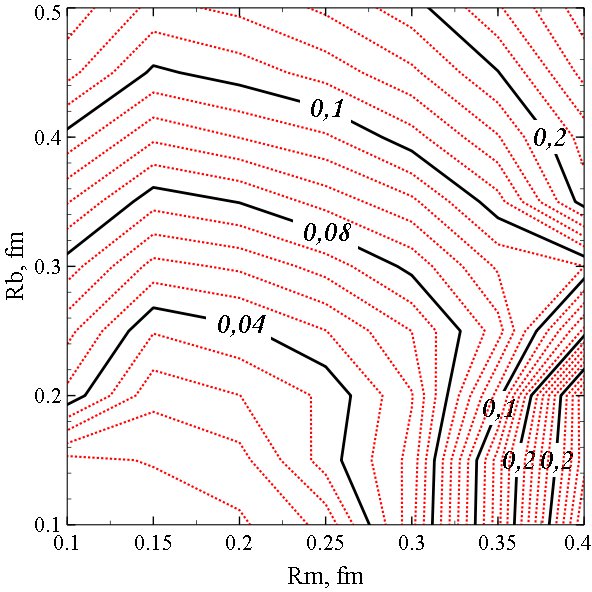}}
 \caption{Difference of $\chi^2/NDF$ between the model without the Lorentz contraction and the model with it for different values of meson and baryon hard-core radii.} \label{chi2_comp}
\end{figure}

 The important  result, however,  is that  the Lorentz contraction inclusion  provides  us with the better fit quality for  any pair of radii $(R_m, R_b)$. From Fig.  \ref{chi2_comp} one can see the difference between the  $\chi^2/NDF$ values found  without  the Lorentz  contraction  and with  it. Obviously, when both radii are small, then the correction due to the  Lorentz contraction is small too. At $(R_b, R_m)$ = (0.3, 0.4) fm $\Delta\chi^2/NDF \approx 0.1$, while $\chi^2/NDF$ itself is 1.48.

\begin{figure}[htbp]
  \centerline{
    \includegraphics[height=7cm]{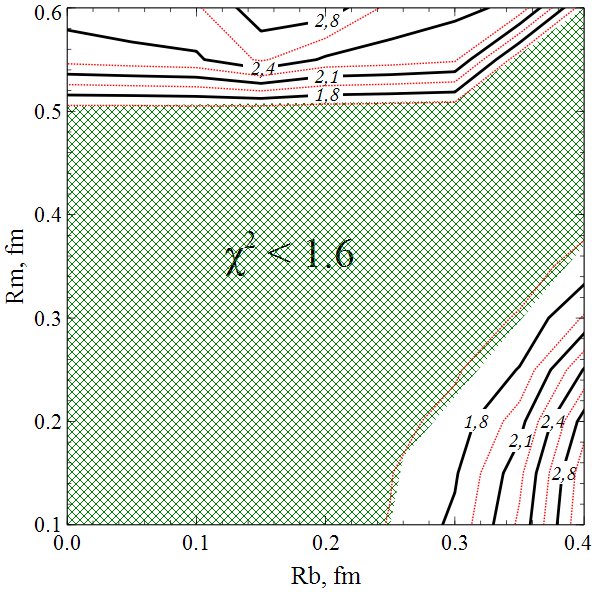}
  }
    \centerline{
   \includegraphics[height=7 cm]{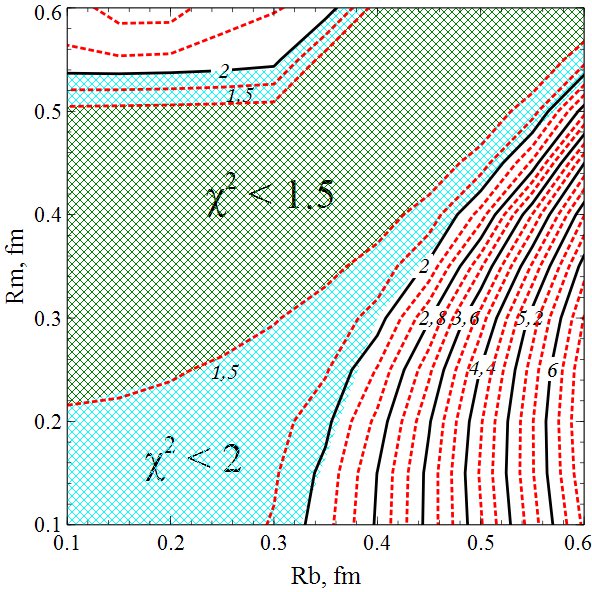}
  }
 \caption{Upper panel:  $\chi^2/NDF$ for the model without the Lorentz contraction for different values of meson and baryon hard-core radii.
Lower panel: same as in the upper panel, but  for the model with the Lorentz contraction included.}
  \label{chi2_contour}
\end{figure}

\begin{table*}
\begin{tabular}{|c|c|c|c|c|}
\hline
Collision energies set, $\sqrt{S_{NN}}$ & $\chi^2$/NDF without surface tension & $\chi^2$/NDF with surface tension & $\sigma_0, {\rm MeV\,fm}^{-2}$ & $T_0$, MeV \\
\hline
2.7 - 7.6 & 25.8135/33 = 0.782 & 25.8043/31 = 0.832 & $0.91\cdot 10^{-2}$ & 61 \\
\hline
2.7 - 200 & 103.096/82 = 1.2573 & 103.036/80 = 1.288 & $-1.37\cdot 10^{-2}$ & 57 \\
\hline
2.7 - 62.4 (no 130 and 200) & 85.51/65 = 1.3156 & 85.268/63 = 1.3534 & $-3.21\cdot 10^{-2}$ & 62 \\
\hline
12, 17, 62.4, 130, 200 & 62.5452/37 = 1.69 & 62.1454/35 = 1.776 & 0.654 & 147 \\
\hline
\end{tabular}
\vspace{0.5cm}
\caption {Results of the global fit, including the extracted surface tension parameters.}
\label{tab:surf_ten}
\end {table*}

The simplest way to obtain the best fit radii is to perform a global fit, including the radii into the fitting procedure. However, in the $(R_m, R_b)$ plane there exist the domains, where $\chi^2$ stays almost unchanged. For example, in the case without the Lorentz contraction $\chi^2$ is the same along the line $R_b=R_m$, which follows directly from (\ref{EqXIX}). This makes the straightforward global fit rather difficult. Therefore,  we perform the fit procedure of particle ratios in central nucleus-nucleus  collisions at $\sqrt{S_{NN}}$ = 2.7, 3.3, 3.8, 4.3, 4.9, 6.3, 7.6, 8.8, 12, 17, 130, 200 GeV  for each pair of the  radii $(R_m, R_b)$ and find the domains,  where $\chi^2$ differs from its minimal value less than 10\%. The results are shown in Fig.  \ref{chi2_contour}.

\section{Determination of  hadronic surface tension}

Recently the extremely important role of  the surface tension of  quark gluon bags
was realized  within the exactly solvable models for the deconfinement phase transition with the tricritical \cite{Bugaev:TriCEP,Aleksei:TriCEP} and the critical \cite{Bugaev:CEP} endpoints.  It was shown \cite{Bugaev:TriCEP, Aleksei:TriCEP, Bugaev:CEP} that
the (tri)critical endpoint appears due to vanishing surface tension coefficient, while
at low baryonic densities
the deconfinement phase transition degenerates into a cross-over just due to the negative values of surface tension coefficient. The existence of  negative values
of the surface tension coefficient at the cross-over temperature for demonstrated
analytically within the model of color confining tube \cite{Bugaev:ColorTube}.
Using this model it was possible to predict the value of (tri)critical temperature
of QCD phase diagram
$T_{cep} = 152.9 \pm 4.5 $ MeV \cite{Bugaev:CepT} using the plausible assumption on the temperature dependence of the  surface tension coefficient  $\sigma (T) = a^2 \left(1 - \frac{T}{T_{cep}}\right)$  \cite{Fisher_67}
which is typical
for ordinary liquids.

Since the lattice QCD is not reliable at the non-zero values of the baryonic chemical
potential it would be interesting to study
the surface tension for hadrons at the chemical freeze-out. The surface free energy  can be written  as $F_{surf}=\sigma(T) S$, where the hadron surface  $S$ is given by its    hard-core radius $R$ as  $S = 4\pi R^2$.
 Note that such a parameterization of the surface free energy is typical for the multi-component  gas  mixtures  \cite{Bugaev:TriCEP,Aleksei:TriCEP,Bugaev:CEP,SMM}.
Inclusion of  the surface free energy into the thermal model is equivalent to adding  the term $\sigma(T) S$ to the total chemical potential. If  $S$ is the same for all hadrons, such a surface tension correction does not affect the hadron yield  ratios. Therefore,  we have taken $R_m$ = 0.45 fm, $R_b$ = 0.3 fm, to have the noticeable  radii difference. In the actual simulations   the surface tension coefficient $\sigma(T)$ was  parameterized as
\begin{eqnarray}\label{EqXXI}
\sigma(T) = \sigma_0   \left(1-  \frac{T}{T_0}  \right) \,.
\end{eqnarray}
Here $\sigma_0$ and $T_0>0$ are the free parameters to be found  from a global fit.
Note that for $T \le T_0$ such a temperature dependence coincides with the
famous Fisher droplet model parameterization \cite{Fisher_67}, whereas
for $T > T_0$ it is in line with the recent findings   \cite{Bugaev:TriCEP,Aleksei:TriCEP, Bugaev:CEP,Bugaev:ColorTube,Bugaev:CepT}.
We have performed several  global fits with the parameterization (\ref{EqXXI}), using different data sets. The results are listed in the Table~\ref{tab:surf_ten}.

From this  table one can conclude that the inclusion of the surface tension
in the form (\ref{EqXXI}) does not improve the fit quality, however, since in all cases the
value of $\chi^2/NDF$ is almost the same as without accounting for the surface tension, it also does not
spoil the fit quality. This fact allows us to take the obtained values of  parameters  $\sigma_0$ and $T_0$ rather seriously.
It is not surprising, that the value of $\sigma_0$ is close to zero, otherwise the sizable surface tension
of  hadrons could be already found. The  really surprising fact is that  for  the center of mass
energies $\sqrt{S_{NN}} \ge  12$  GeV  the parameter $T_0 = 147 \pm 7$ MeV is extremely
close to the critical temperature value  $T_{cep} = 152.9 \pm 4.5 $ MeV found in \cite{Bugaev:CepT}
more than a year ago using entirely different approach.  Of course, the reason of why the global fit that includes
the  low energy data gives essentially lower value of the parameter $T_0$ should be understood, and, hence,
the investigation of the hadronic surface tension should be continued  using both the experimental data on hadron production and
the  lattice QCD data.

\section{Multi-component hadron gas and the Strangeness Horn description}

  The thorough  analysis performed above led us to a conclusion that besides the hadron surface tension inclusion the further improvement  of the thermal model can be achieved, if we consider the pion and kaon hard-core radii, as an independent fitting parameters. On the one hand this would allow us  to have two additional fitting parameters, and on the other hand one could include the Strangeness Horn data into the fitting procedure. Note that up to now  the quality of the Strangeness Horn description is far from being satisfactory, although very different formulations of  the thermal model are used for this purpose \cite{R_pi1, Andronic_big, Horn:2012}.
Thus, the most recent compilation of the Strangeness Horn description by the different thermal models
can be found in  \cite{Horn:2012}.

To improve the Strangeness Horn description can be easily understood from the fact that just the non-monotonic behavior of the $K^+$/$\pi^+$ ratio as the function of the center of mass energy of collision is often claimed to be one of a few existing signals of the onset on deconfinement \cite{Gazd_Horn:99, Step, Gazd_rev:10}. The multi-component hadron gas model developed
in \cite{Bugaev_two_comp_VdW} is perfectly suited to treat the pion and kaon hard-core radii as independent fitting parameters. The physical idea behind such an approach is that the
hadronic hard-core radii  are the effective parameters which include the contributions of the repulsion and attraction. Since the parameters of hadron-hadron interaction are, generally speaking,  individual for each kind of hadrons, then each kind of hadrons can have  its own
hard-core radius.  Based on this idea we performed a global fit of all hadron multiplicities  as described above  together with the Strangeness Horn data considering the pion hard-core radius $R_\pi$ and the kaon hard-core radius $R_K$ as independent variables together with the chemical freeze-out temperature and baryonic chemical potential, whereas the hard-core radius of all other mesons and the hard-core radius of baryons were  fixed, respectively,  as $R_b = 0.3$ fm and
$R_m = 0.5$ fm according to the findings  of Section 6.

\begin{figure}[htbp]
  \centerline{
    \includegraphics[height=7.7cm]{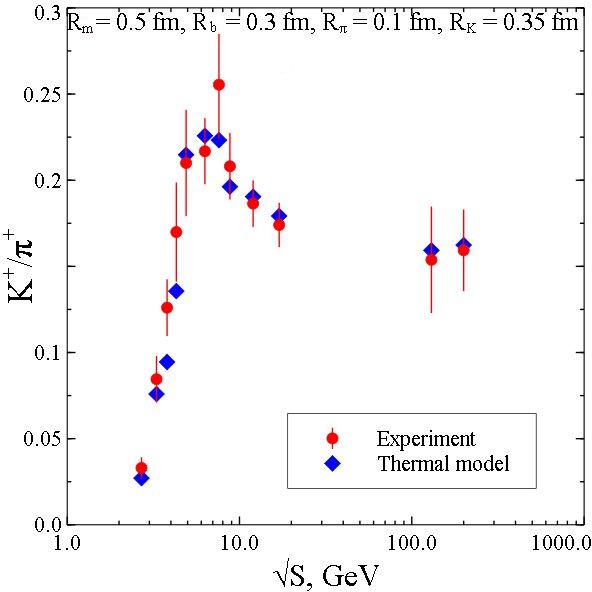}
  }
     \caption{Strangeness Horn description  for the model with pion and kaon hard-core radii to be independent fitting parameters. The resulting quality of the global fit is  $\chi^2/NDF \simeq 1.019$.}
  \label{Horn_new}
\end{figure}

The results of such a fit are shown in Fig. \ref{Horn_new}. Comparing Figs. \ref{str_horn} and
\ref{Horn_new} one can see the dramatic improvement of the  $K^+$/$\pi^+$ ratio for the collision energies $\sqrt{S_{NN}}$ above 8 GeV. In fact,  the $K^+$/$\pi^+$ ratio shown in Fig. \ref{Horn_new} deviates  from the experimental error  bars on for the energy  $\sqrt{S_{NN}} = 3.8$ GeV,
while for all other collision energies it does not  essentially deviate from the experimental error  bars.   Thus, the variations of pion and kaon hard-core radii essentially improve  the description of this
ratio for all collision energies, but at the same time  the quality of the fit of all other hadronic multiplicities  does not worsen as it is seen
from the resulting value of  $\chi^2/NDF \simeq 1.019$ for the global fit, which, so far,  is the best result obtained in the literature.

\section{Conclusions}

\par
In this work we performed a comprehensive analysis of the experimental hadron multiplicities
within the thermal model.
As in previous studies the  considered thermal model has  two hard-core radii ($R_b$ for baryons and $R_m$ for mesons) and two   new elements: an inclusion of the Lorentz contraction of eigen volumes of hadrons and  a treatment of hadronic surface tension.
Using this model we studied the role of  the imposed conservation laws (\ref{EqV}) and (\ref{EqVI}), and showed that for the adequate   description of hadron multiplicities the   conservation laws should be modified, whereas  for the description of  hadron yield ratios  the conservation laws are not necessary at all.
In addition,  we suggested and analyzed  the thermal model in which the pion and kaon hard-core radii are  independent fitting parameters compared to all other mesons.

Here we also analyzed the usual criteria for the chemical freeze-out and found that none of them is robust.
Therefore we suggested a novel criterion of chemical freeze-out, a constant value of entropy per hadron number equals to  7.18.
Such a criterion is also  supported by the different formulation of thermal model  \cite{Andronic_big} and it evidences for the new physical effect which we called the {\bf adiabatic chemical  hadron production}.

The performed analysis allowed us to find the restrictions  on the hard-core radii that are imposed by the experimental data. Also we  showed  that although an inclusion of the  Lorentz contraction improves  the fit quality for any pair of baryon and meson hard-core radii, but at the same time it has a small effect on the chemical freeze-out parameters and on  $K^+/\pi^+$ ratio.

In addition we phenomenologically  introduced the  surface tension in thermal model and made several global fits to find its parameters.
Although, the surface tension inclusion does not improve the fit quality,  for the first time  it is  found that  the temperature of  the nil surface tension value depends on the considered interval of the collision energy. Thus, if the low energy data are included into the fit, then the nil surface tension temperature is about $60 \pm 5$  MeV, while the data
for the center of mass  energies above 10 GeV lead to an essentially larger value of  this temperature value $T_0 = 147 \pm 7$ MeV. The latter is a very intriguing result since a very close estimate for the   nil surface tension temperature was obtained recently within entirely different analysis of  the quark gluon bag  surface tension  \cite{Bugaev:ColorTube}. Therefore, it is possible that
these two independently obtained results, indeed, evidence that the (tri)critical temperature of the
QCD phase diagram is between 140 and 154 MeV.

The most dramatic numerical  effect, however, is obtained for the truly multi-component hadron gas model worked out  in \cite{Bugaev_two_comp_VdW} and employed  here for the first time. In this model  the hard-core radii of pions and kaons differ from the hard-core radius  of all other mesons and they   are treated as independent fitting parameters. Such an approach allowed us
for the first time to simultaneously  fit   the hadron multiplicities together with the Strangeness Horn and to get   the  chemical freeze-out data description  of very high quality.

\section{Appendix: Heuristic derivation of the approximate excluded volume formula for ellipsoids of revolution}

In order to fasten the numerical evaluation of the relativistic excluded volumes we would like to obtain
an approximate expression which would reduce the dimension of momentum integrations  of two particles from
six to three. Basically here we employ the heuristic method suggested in \cite{Bugaev_Lorentz_cont_2}.
The main difference, however, is that in \cite{Bugaev_Lorentz_cont_2} the ultra-relativistic expression for the excluded volume  was
derived, while here we would like to get an expression which would be accurate both in the non-relativistic and
ultra-relativistic  limits.

\par
\begin{figure}[htbp]
    \includegraphics[height=6cm]{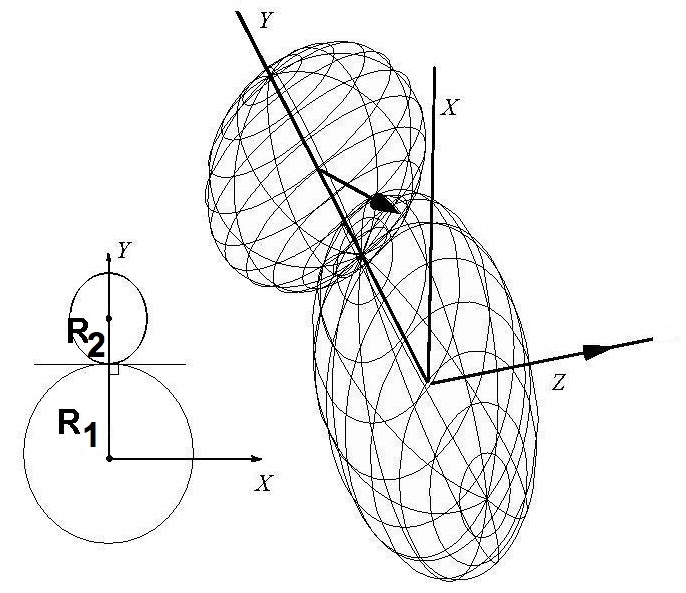}
   \caption{Explanation on how  to obtain the expression for  the  radius $R_y$ of the relativistic  excluded volume, when the second ellipsoid is translated around the first one in the plane XOY.}
    \label{Y_ellipsoids}
\end{figure}

\begin{figure}[htbp]
    \includegraphics[height=6cm]{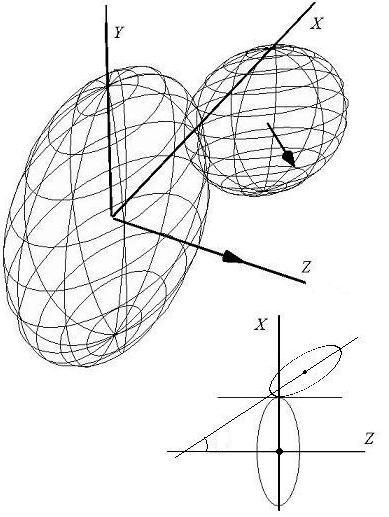}
   \caption{Explanation on how  to obtain the expression for  the  radius $R_x$  of the relativistic  excluded volume, when the second ellipsoid is translated around the first one in the plane XOZ.
}
    \label{X_ellipsoids}
\end{figure}

  For this purpose let us consider two relativistic spheres $S_1$ and $S_2$, there $\gamma$-factors being $\gamma_1$ and $\gamma_2$, respectively.
The hard-core radii in their rest frames are $R_1$ and $R_2$, respectively. Let us fix an angle $\theta$ between the  momenta  of  two particles   which is the standard spherical angle. It  is chosen in such a way that
 the three momentum of $S_1$ coincides with $OZ$-axis  (see Figs. \ref{Y_ellipsoids} and \ref{X_ellipsoids} for details).
 Then the angle $\theta$ is the azimuthal spherical angle of the momentum of second particle.
 Due  to the Lorentz contraction the both spheres shrink in the direction of their momenta and  one obtains  two ellipsoids of revolution. Here we show how to get  an approximate formula for the excluded volume for such ellipsoids.

\par The  basic idea is to neglect the complexity of the problem and treat the excluded volume as an ellipsoid, afterwards  to symmetrize the obtained  expression with respect to interchange $1 \leftrightarrow 2$ and take the half of the sum of  two expressions.
Then, the unsymmetrized  excluded volume reads:
\begin{eqnarray}\label{EqXXII}
V_{exc} = \frac{4}{3}\pi R_x R_y R_z \,,
\end{eqnarray}
where the ellipsoid's radii  $R_x$, $R_y$, $R_z$ are found from geometrical consideration given below.

\begin{figure}[htbp]
    \includegraphics[height=6cm]{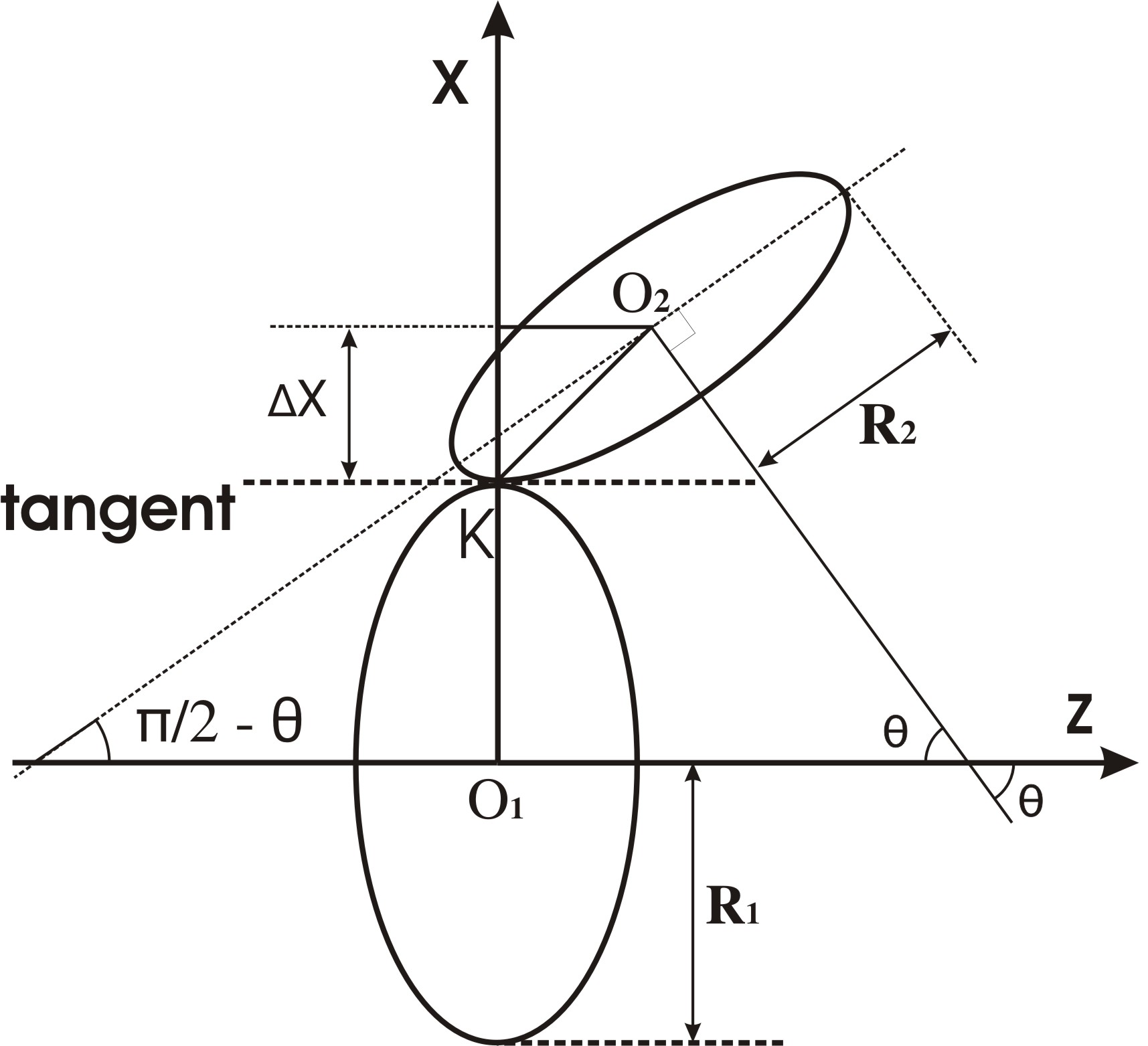}
   \caption{The detailed projection of the second ellipsoid translation around the first one in  the plane XOZ. This is an explanation on how to derive $R_x$ from Fig. \ref{X_ellipsoids}. }
    \label{XZ_flat}
\end{figure}

\begin{figure}[htbp]
    \includegraphics[height=6cm]{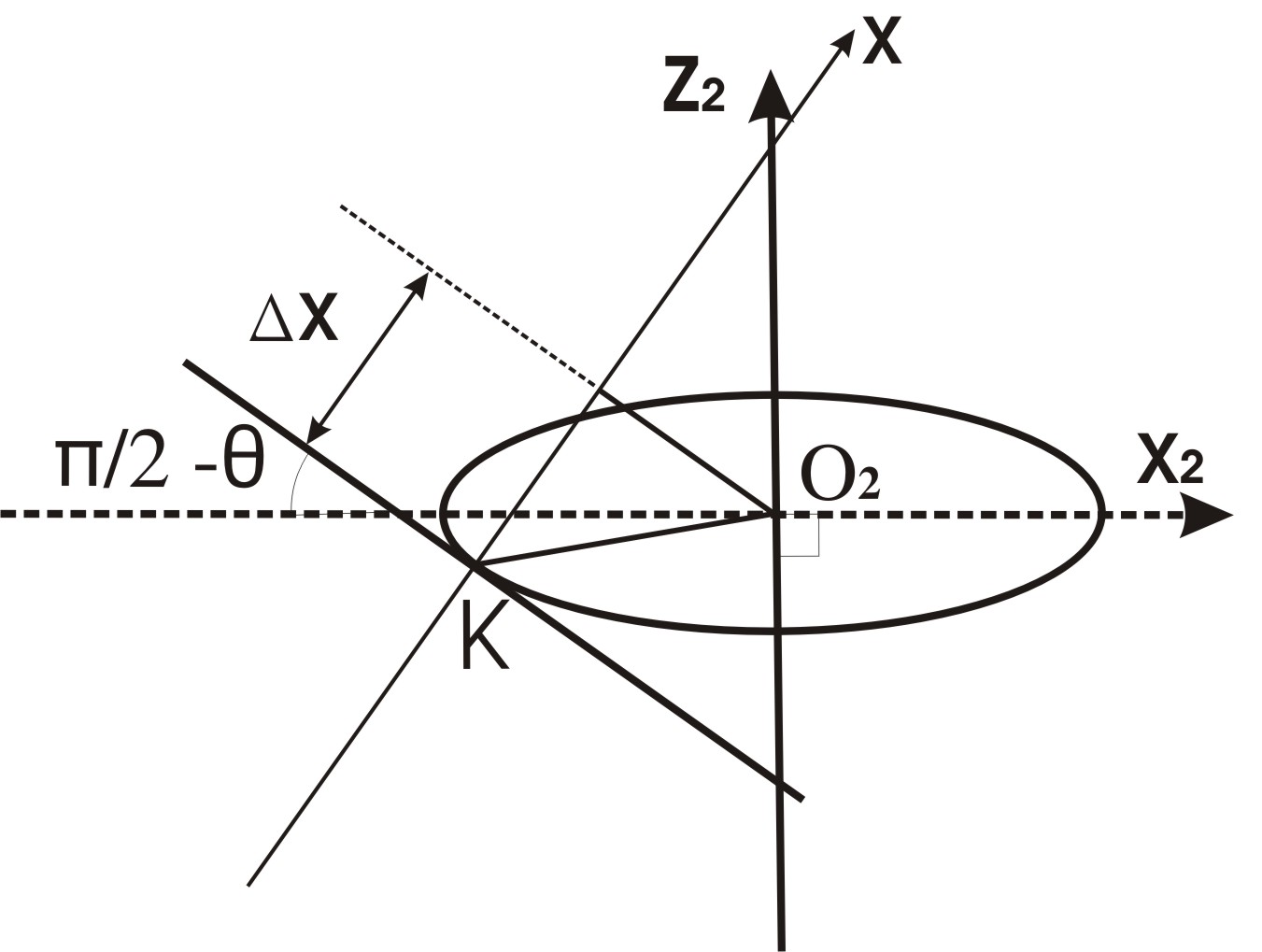}
   \caption{The fragment of  Fig. \ref{XZ_flat} which is necessary to determine the coordinates of the touching point K in the coordinate system of  $S_2$.}
    \label{XZ_main_axes}
\end{figure}

\begin{figure}[htbp]
    \includegraphics[height=6cm]{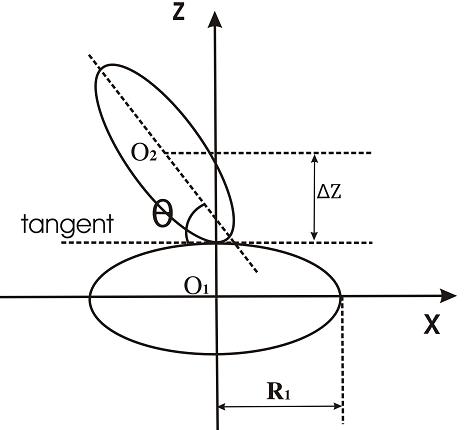}
   \caption{The projection which is necessary to derive the radius  $R_z$ of the approximate excluded volume (\ref{EqXXXIII}).
   Comparing this projection with that one shown in Fig. \ref{XZ_flat}, we find  that formally it is necessary to replace $R_1 \rightarrow R_1/\gamma_1$ and $\pi/2 - \theta \rightarrow \theta$ in the expression  (\ref{EqXXIV}) to get the  expression for $R_z$ from that one for  $R_x$.}
    \label{ZX}
\end{figure}

From Fig. \ref{Y_ellipsoids} one can see  that
\begin{eqnarray}\label{EqXXIII}
R_y = R_1 +R_2\, .
\end{eqnarray}
To obtain the radius  $R_x$ one should consider the ellipsoids  depicted in Figs. \ref{X_ellipsoids}--\ref{XZ_main_axes}. The radius  $R_z$ can be found analogously to the radius $R_x$ from  Fig. \ref{ZX}.

Let us show how one can get an  expression for  the radius $R_x$. A convenient projection and the notations are depicted  in Figs. \ref{X_ellipsoids} and \ref{XZ_flat}. From Fig.  \ref{XZ_flat} one gets

\begin{eqnarray}
R_x=R_1+\Delta x \,.
\end{eqnarray}

Turning the reference frame to the main axes $x_2$ and $y_2$ of $S_2$ (see fig. \ref{XZ_main_axes}) one easily finds the coordinates of touching point $K(x_0,y_0)$. The equation of $S_2$ for  the principal axes
shown  in Fig. \ref{XZ_main_axes} reads as:
\begin{eqnarray}\label{EqEllips}
x_2^2+y_2^2\gamma_2^2=R_2^2 \,.
\end{eqnarray}
The equation for a  tangent to an ellipse shown   in Fig. \ref{XZ_main_axes} is:
\begin{eqnarray}\label{EqTangent}
\frac{dy_2}{dx_2} = -\frac{x_2}{\gamma_2^2 y_2} = \tg(\pi/2 - \theta) \,.
\end{eqnarray}

Now solving (\ref{EqEllips}) together with (\ref{EqTangent}) one gets the  touching  point coordinates $x_0$ and $y_0$:
\begin{eqnarray}
 x_0 &= -\frac{R_2\gamma_2 \ctg\theta}{\sqrt{1+\gamma_2^2 \ctg^2\theta}} \, ,\\
y_0 &= \frac{R_2}{\gamma_2\sqrt{1+\gamma_2^2 \ctg^2\theta}} \, .
\end{eqnarray}
Turning back from ($x_2$,$y_2$)-coordinate system  to ($x$,$y$)-system,  in  accord with  Figs. \ref{XZ_flat} and \ref{XZ_main_axes} one obtains
\begin{eqnarray}
\Delta x = |x_0 \cos\theta - y_0 \sin\theta| =\frac{R_2 \sin\theta}{\gamma_2} \sqrt{1+\gamma_2^2 \ctg^2\theta}\,,
\end{eqnarray}
and hence we get
\begin{eqnarray}\label{EqXXIV}
R_x=R_1+\frac{R_2 \sin\theta}{\gamma_2} \sqrt{1+\gamma_2^2 \ctg^2\theta} \, .
\end{eqnarray}
From Fig. \ref{ZX} one can see that in order to obtain the radius  $R_z$ one should simply  replace $R_1 \rightarrow R_1/\gamma_1$ and $\pi/2 - \theta \rightarrow \theta$ in expression  (\ref{EqXXIV}) for the radius $R_x$. Then one finds
\begin{eqnarray}\label{EqXXV}
R_z =\frac{R_1}{\gamma_1}+\frac{R_2 \cos\theta}{\gamma_2} \sqrt{1+\gamma_2^2 \tg^2\theta} \,.
\end{eqnarray}

Finally, substituting Eqs.  (\ref{EqXXIII}), (\ref{EqXXIV}) and  (\ref{EqXXV}) into (\ref{EqXXII}) one obtains:
\begin{eqnarray}\label{EqXXXII}
V_{exc} & = & \frac{4\pi}{3} (R_1+R_2)\left(\frac{R_1}{\gamma_1}+\frac{R_2 \cos\theta}{\gamma_2} \sqrt{1+\gamma_2^2 \tg^2\theta}\right)  \nonumber \\
&\times & \left(R_1+\frac{R_2 \sin\theta}{\gamma_2} \sqrt{1+\gamma_2^2 \ctg^2\theta}\right) \,.
\end{eqnarray}
Evidently, the expression above  precisely recovers   the excluded volume of two non-relativistic spheres,
 i.e. for $\gamma_1=\gamma_2 = 1$. Thus, in contrast to the result of  \cite{Bugaev_Lorentz_cont_2} Eq. (\ref{EqXXXII}) gives an exact result for the non-relativistic particles.
Also one can analytically  show that Eq. (\ref{EqXXXII})  gives rather good approximation (with the deviation below 10 \% from the exact result)  also for other extreme cases, when one particle is non-relativistic and another particle is ultra-relativistic or when both particles are ultra-relativistic.

Nevertheless, in order to improve  (\ref{EqXXXII}) further, we  symmetrize it with respect to interchange $1 \leftrightarrow 2$. Evidently,  such a procedure will not break down the above discussed properties of
 (\ref{EqXXXII})
However, a symmetrization of  (\ref{EqXXXII})  can be done in many possible ways, but   numerically we found that the best approximation to an exact result
is given by the expression
\begin{eqnarray}\label{EqXXXIII}
V^{rel}_{exc} &=& \frac{\pi}{12}(R_z+\tilde{R}_z)\left[(R_x+R_y)^2+(\tilde{R}_x+\tilde{R}_y )^2\right] \,,
\end{eqnarray}
where  the tilded values $\tilde{R}_a(R_1,R_2,\gamma_1,\gamma_2) = R_a(R_2,R_1,\gamma_2,\gamma_1)$ stand for an  interchange $1 \leftrightarrow 2$ in the expressions for  the radii $R_x$, $R_y$ and $R_z$ of this Appendix. An expression (\ref{EqXXXIII}) leads to   the following  approximate value of the second virial coefficient \cite{Bugaev_Lorentz_cont_2}
\begin{eqnarray}\label{EqXXXIV}
V_{rel} &=& \frac{1}{\rho(T,m_1) \rho(T,m_2)} \int \frac{d^3k_1}{(2 \pi)^3} \frac{d^3k_2}{(2 \pi)^3} \,
 V^{rel}_{exc} (k_1,k_2, \Theta_2)
\nonumber \\
 &\times& \exp{\left( - \frac{\sqrt{k_1^2 + m_1^2}}{T} \right)} \, \exp{\left( - \frac{\sqrt{k_2^2 + m_2^2}}{T} \right)}\, ,
\end{eqnarray}
which, evidently, can be analytically integrated over three spherical angles.
In  Eq. (\ref{EqXXXIV}) the thermal density of the particle of mass $m$ at temperature $T$ is defined as
\begin{eqnarray}\label{EqXXXV}
\rho(T,m) &=&  \int \frac{d^3k_1}{(2 \pi)^3}
 \exp{\left( - \frac{\sqrt{k_1^2 + m^2}}{T} \right)} \, .
\end{eqnarray}

\begin{figure}[htbp]
    \includegraphics[height=7.2cm]{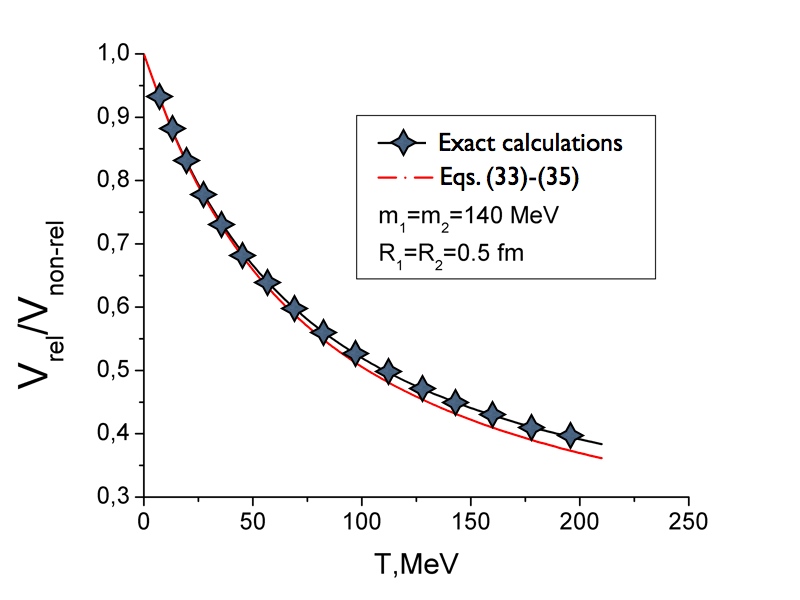}
   \caption{Temperature dependence of  the exact second virial coefficient of two pions (curve with symbols)  and its approximation (curve without symbols) given by
   Eqs. (\ref{EqXXXIII})-(\ref{EqXXXV}) in the units of non-relativistic excluded volume $V_{non-rel}= \frac{32}{3}\pi R_1^3$ of two hard spheres of the same radius $R_1=R_2 = 0.5$ fm.
The comparison is made for pion-pion interaction only, since  in this case the effect of Lorentz contraction is strongest.}
    \label{V_d_compar}
\end{figure}

A comparison  between the exact  value of  the second virial coefficient  with the Lorentz contraction accounted for both particles  and the second virial coefficient  found from  the  approximation  (\ref{EqXXXIII})  is depicted in Fig. \ref{V_d_compar}.
For such a comparison we choose the worst possible case, i.e.  the largest allowed  values of  the hard-core radii and take the pions since for them the relativistic effects are most important.  As one can see from Fig. \ref{V_d_compar}
the approximate expression (\ref{EqXXXIII}) gives very good description of the pion-pion second virial coefficient. In fact,  for temperatures below 180 MeV the relative deviation of the obtained  approximation
 does not exceed 6\%. Since the hard-core repulsion provides a small correction (less than 10\%)  to the system pressure, then the resulting error for the pion pressure generated by  the  approximation  (\ref{EqXXXIII})  is less than 0.5\% for all considered temperatures. The correction to the pressure of heavier hadrons is practically negligible, since the relativistic effects for them are essentially weaker than for pions.

\vskip3mm

{\bf Acknowledgments.}
We would like to thank A. Andronic for  providing an access to well-structured experimental data
and A.I. Ivanytskyi, I.N. Mishustin and L.M. Satarov  for fruitful discussions.
K.A.B. and D.R.O.  acknowledge  the partial  support of the Program `On Perspective Fundamental Research in High Energy and Nuclear Physics' launched by the Section of Nuclear Physics  of National Academy of  Sciences of Ukraine.
The work of  A.S.S.  was supported in part by the Russian
Foundation for Basic Research, Grant No. 11-02-01538-a.

\end{document}